\documentstyle{mn}
\newcommand{\bbu}{{\bmath u}}
\newcommand{\bbn}{{\bmath n}}
\newcommand{\bbe}{{\bmath e}}
\begin{document}

\title[Incompressible fluid binary]
{Incompressible fluid binary systems with internal flows -- 
models of close binary neutron star systems with spin}
\author[K. Ury\=u and Y. Eriguchi]
{K. Ury\=u$^{1,2}$ and Y. Eriguchi$^3$\\
$^1$International Center for Theoretical Physics, Strada Costiera 11,
34100 Trieste, Italy\\
$^2$SISSA, Via Beirut 2/4, 34013 Trieste, Italy\\
$^3$Department of Earth Science and Astronomy, 
Graduate School of Arts and Sciences, 
University of Tokyo, Komaba, Meguro, Tokyo, Japan
}

\maketitle

%
%%%%%%%%EEEEEEEEEEEEEEEE
\begin{abstract}
In this paper we have examined {\it numerically exact} configurations of close 
binary systems composed of incompressible fluids with internal flows.  
Component stars of binary systems are assumed to be circularly orbiting each 
other but rotating nonsynchronously with the orbital motion, i.e. stars in 
binary systems have steady motions seen from a rotational frame of reference.  
We have computed several equilibrium sequences by taking fully into account 
the tidal effect of Newtonian gravity without approximation.  We consider two
binary systems consisting of either 1) a point mass and a fluid star or 2)
a fluid star and a fluid star.  Each of them corresponds to generalization 
of the Roche--Riemann or the Darwin--Riemann problem, respectively.  
Our code can be applied to various types of incompressible binary systems 
with various mass ratios and spin as long as the vorticity is constant.  

We compare these equilibrium sequences of binaries with approximate solutions
which were studied extensively by Lai, Rasio and Shapiro (LRS) as models for 
neutron star--neutron star (NS--NS) binary systems or black hole--neutron star 
(BH--NS) binary systems.  Our results coincide qualitatively with those of 
LRS but are different from theirs for configurations with small separations.  
For these binary systems, our sequences show that dynamical or secular 
instability sets in as the separation decreases.  The quantitative errors of 
the ellipsoidal approximation amount to $2\sim 10\%$ for configurations
near the instability point.  Compared to the results of LRS, the separation 
of the stars at the point where the instability sets in is larger and 
correspondingly the orbital frequency at the critical state is smaller
for our models.  Since these sequences can be considered as 
evolutionary models of binary NS systems where component stars are 
approaching due to the effect of gravitational wave emission, we can expect 
that the final fate of such binary systems will be dynamical coalescence 
due to dynamical instability.  
%%%%EEEEEEEEEEEEEEE
\end{abstract}

%%%%EEEEEEEEEEEEEEE

\section{Introduction}

Close binary systems composed of massive compact objects such as neutron 
stars or black holes are relevant systems in astrophysics, because they 
are possible sources of many varieties of astrophysical phenomena.  A close 
binary system of compact objects will evolve quasi-statically by decreasing 
its orbital separation due to gravitational radiation emission until a 
final stage such as a violent merging or mass overflow will be reached.  
During inspiraling and coalescence phases, binary systems may be observed as 
sources of not only cosmological $\gamma$-ray bursts (see e.g. Paczynski 1986)
but also gravitational waves which will be detectable by kilometer size laser 
interferometers such as LIGO and VIRGO (see e.g. Abramovici et al. 1992; 
Thorne 1994).  

The evolution of binary systems prior to merging was discussed by Kochanek 
\shortcite{ko92} and Bildsten \& Cutler \shortcite{bc92}.  In their papers, 
it was pointed out that, for most viscosities known thus far, the time scale of 
angular momentum transfer within a neutron star is longer than that of 
evolution due to gravitational radiation reaction.  Therefore the effect of 
viscosity may be negligible during the evolution of binary systems toward 
coalescence. In other words, viscosity may not be powerful enough to 
synchronize spin motions nor change spin of component stars.  

In the framework of Newtonian gravity without viscosity, circulation of 
a fluid remains constant even when the gravitational radiation reaction 
potential is included \cite{mi74}.  Therefore, stars in binary systems will 
not rotate synchronously but have internal flows or spin in the frame of 
reference rotating with the orbital angular velocity.  

Because of a long time scale of energy dissipation due to gravitational 
radiation, such evolutionary sequences of binary systems can be followed 
by using equilibrium configurations even until stars come near to the 
innermost stable orbit \cite{st83}.  However, it is very difficult to 
construct stationary states of compressible stars with internal motions even 
today for non-axisymmetric configurations such as binary stars (Ury\=u \& 
Eriguchi 1996, and references therein) except for the case of 
irrotational gaseous star binary systems (Ury\=u \& Eriguchi 1998a, 1998b).  
Several years ago, Lai Rasio \& Shapiro (hereafter LRS) developed a 
semi-analytical method 
to treat equilibrium of incompressible and compressible stars under the 
assumption that stellar shapes and equi-density contours are approximated 
by ellipsoids \cite{lrs93a} (LRS1 hereafter).  In a series of papers, 
they applied their method to binary neutron star systems with spin and 
extensively examined the innermost stable circular orbits where dynamical 
instability of binary systems sets in (Lai, Rasio \& Shapiro 1994a and 1994b: 
hereafter LRS2 and LRS3, respectively).  

Equilibrium configurations of incompressible binary fluid star systems 
with internal motions were computed numerically exactly by Eriguchi and 
Hachisu (1985).  Furthermore, Eriguchi, Futamase and Hachisu (1990) computed 
sequences of binary disks with internal motions.  In these papers they 
computed sequences of equal mass binary disks or stars which can be used
to study evolutionary sequences due to gravitational wave emission by keeping
the circulation of each component constant and showed that there exist turning 
points on equilibrium sequences. In other words, they indicated occurrences 
of secular and dynamical instabilities prior to the work of Lai, Rasio \& 
Shapiro (LRS1).  

In this paper we have extended the computational method developed by Eriguchi 
and Hachisu (1985) so as to compute Roche type binary configurations and/or 
binary systems with arbitrary mass ratios and internal motions, and have 
reexamined equilibrium models of incompressible binaries whose parameters 
are in the physically meaningful range.  First, we have computed sequences 
of {\it synchronously rotating} binary systems consisting of a point 
mass and a fluid, i.e. Roche type binary systems, and have compared our 
results with those obtained by using the ellipsoidal approximation by LRS.  
Second, we have concentrated on configurations of irrotational Roche--Riemann 
and Darwin--Riemann problems for which vorticity seen from the inertial 
frame vanishes.  We have compared our results again with ellipsoidal 
models computed in LRS1 and LRS2.  Our results coincide qualitatively 
with those of LRS.  For these binary systems, our sequences show 
that dynamical or secular instability sets in.  The quantitative errors of 
the ellipsoidal approximation are usually $2\sim 10\%$ for models near the 
instability point.  These sequences of equilibrium binary stars can be used 
as quasi-evolutionary models of binary NS systems in which the separation 
of two component stars is decreasing due to the effect of gravitational 
wave emission.  Consequently such binary systems will result in coalescence 
due to dynamical instability.  

In the next section we present the basic equations and explain briefly the 
numerical method for incompressible fluid binary systems with spin, in 
section 3 we show computational results of binary sequences for several 
types, and in section 4 we discuss evolution of binary systems of compact 
objects by applying our results.   

\section{Formulation of the problem} 

\subsection{Assumptions and basic equations}

We assume plane symmetry 
of stellar structures about the $x$-$y$ plane and the $x$-$z$ plane in the 
rotational frame of reference. Here the Cartesian coordinates $(x,y,z)$ are 
used. The $z$-axis is chosen along the axis of rotation and centers of 
mass of two stars are located on the $x$-axis.  In this paper we consider 
neutron stars which are approximated by {\it incompressible fluids}.  
Although this is a crude approximation, we have not found appropriate
schemes to treat compressible self-gravitating gases with generic steady 
internal flows as mentioned in Introduction (see also Ury\=u \&
Eriguchi 1996).  In particular, it is difficult to formulate a well-posed 
problem and derive proper basic equations for steady states of 
{\it compressible fluids}, when the flow is parallel to the equatorial 
plane (the planar flow) and the vorticity vector $\bzeta$ of the 
fluid (seen from the rotational frame of reference) is parallel to 
the rotational axis, $\bzeta \, = \, \zeta \, \bbe_z$ where $\bbe_i$ 
($i \, = \, x,y,z$) denotes the unit basis vector along each coordinate 
axis.  
%%%%%%%%%%%%%%%%%  begin  %%%%%%%%%%%%%%%%%
These assumptions are not only physically natural, since the 
gravitational radiation reaction potential conserves the circulation of 
the flow field (see below), but also necessary to make the problem 
tractable, since it reduces the number of physical variables.  However, 
for compressible models, we cannot write down the basic equations 
which are consistent with the number of independent physical variables.  
This occurs because the flow becomes independent of $z$ under the above 
assumptions \footnote{However, we can formulate and solve the problem of 
compressible fluid binary consistently for irrotational case, i.e. the 
vorticity of fluid vanishes identically in the inertial frame of reference 
\cite{ue98b}.  In this case, we do not need to require the planar flow 
explicitly.} (see also the formulation for incompressible case below).  
%%%%%%%%%%%%%%%%%   end   %%%%%%%%%%%%%%%%%

On the other hand, for {\it incompressible fluids} we can formulate the
problem properly by introducing the stream function for planar flows as 
follows: 
\begin{equation} \label{strea}
u_x \, = \,  {\partial\,\Psi \over\,\partial y} \, , \quad
u_y \, = \, -{\partial\,\Psi \over\,\partial x} \, ,
\end{equation}
where $\Psi$ and $(u_x, \, u_y)$ are the stream function and $x$- and
$y$-components of the velocity vector in the rotational frame, 
respectively.  Note that $\Psi$ and $(u_x, \, u_y)$ depend only on $(x,y)$ 
and that the equation of continuity is automatically satisfied.
Thus we can write down basic equations and boundary conditions consistently.  

If the stream function is a function of the $z$-component of vorticity 
vector, $\zeta$, which we will refer to as the vorticity, 
the Euler equation of fluid motion can be integrated to the Bernoulli 
equation.  Since we consider the case with $\zeta = 
{\rm const}$ everywhere in the fluid, the Bernoulli equation becomes as 
follows: 
\begin{eqnarray} \label{berno}
\lefteqn{{1 \over 2}(u_x^2 + u_y^2)\,+\,(\zeta\,+\,2\Omega)\Psi}
\nonumber \\
&&\qquad\qquad -\,{1 \over 2}\Omega^2(X^2+Y^2)
+\,{p \over \rho}\,+\,\phi\,= \, C \, ,
\end{eqnarray}
where $\Omega$, $p$, $\rho$, $\phi$ and $C$ are the orbital angular velocity, 
the pressure, the density, the Newtonian gravitational potential and the 
Bernoulli constant, respectively. The quantity $X^2+Y^2$ means the square of 
the distance of fluid element from the rotational axis.  

By substituting the velocity components Eq.~(1) into the following
definition of vorticity 
\begin{equation}
\zeta\,=\,{\partial\,u_y \over \partial\,x}\,-\,
          {\partial\,u_x \over \partial\,y}\, ,
\end{equation}
the equation for the stream function can be derived as, 
\begin{equation} \label{stre2}
\triangle\Psi\,= \, - \zeta \, , 
\end{equation}
where 
\begin{equation} 
\triangle \, = \, {\partial^2 \over \partial x^2}\,+\,
                     {\partial^2 \over \partial y^2}\, .
\end{equation}

As for the gravitational potential, we use the integral form of the Poisson 
equation as follows:
\begin{equation} \label{pot}
\phi({\bf r}) = - \,G \rho\int_V {1\over \mid {\bf
r-r'} \mid } \, d^3 {\bf r'} \, .
\end{equation}
Here $G$ is the gravitational constant and $V$ is the total volume over 
which the fluid extends.

Boundary conditions for the equations described above are as follows: 
\begin{equation} \label{bcon1}
p(x,y,z)  = 0, \quad \,
{\rm on\ the\ surface\ of\ the\ star}, 
\end{equation}
and
\begin{eqnarray} \label{bcon2}
\lefteqn{\Psi(x,y) = A,}  \nonumber \\ 
&&\quad {\rm along\ the\ surface\ on\ the\ equator\ of\ the\ star}, 
\end{eqnarray}
where $A$ is a constant.  
%%%%%%%%%%%%%  begin  %%%%%%%%%%%%%%%%%%%%%
Since we 
%%%%%%%%%%%%%  end  %%%%%%%%%%%%%%%%%%%%%
can choose its value arbitrary,
%%%%%%%%%%%%%  begin  %%%%%%%%%%%%%%%%%%%%%
we 
%%%%%%%%%%%%%   end   %%%%%%%%%%%%%%%%%%%%%
set $A=0$ for convenience.  It should be noted that boundary conditions for 
the gravitational potential are already included in the integral 
representation Eq.~(\ref{pot}).  

\subsection{Solving method}
In actual computations, we introduce two spherical coordinate systems 
$(r_\pm, \theta_\pm, \varphi_\pm)$ whose origins are fixed on the $x$-axis at
distances
\begin{equation} \label{dist}
d_\pm \, \equiv 
\, {\mid R_\pm^{\rm out} \, + \, R_\pm^{\rm in} \mid \over 2} \ ,
\end{equation}
for each fluid component, where $R_\pm^{\rm out}$ and $R_\pm^{\rm in}$ are 
distances from the rotational axis to the outer and inner edges of the star
on the $x$-axis, respectively.  Here subscripts $+$ and $-$ correspond to 
the primary and the secondary stars, respectively.  For Roche type binary 
systems, the primary star corresponds to a fluid star and the secondary 
to a point mass.  Therefore note that usually the mass of primary $M_+$ is 
smaller than that of secondary $M_-$, i.e. $M_+\le M_-$.  
The angle $\theta_\pm$ is the zenithal angle measured from the positive 
direction parallel to the $z$-axis and the angle $\varphi_\pm$ is the 
azimuthal angle measured from the positive direction of the $x$-axis.  
Relations between the Cartesian coordinates
$(X_\pm,Y_\pm,Z_\pm)$ and the polar coordinate systems 
$(r_\pm, \theta_\pm, \varphi_\pm)$ are expressed as follows:
\begin{eqnarray}
X_\pm\,&=&\,\pm d_\pm \,+ \, r_\pm \, \sin \theta_\pm \, \cos \varphi_\pm \, , 
\nonumber\\
Y_\pm\,&=&\,\pm d_\pm \,+\,r_\pm \, \sin \theta_\pm \,\sin \varphi_\pm \, , \\
Z_\pm\,&=&\,r_\pm \, \cos \theta_\pm \, . 
\nonumber
\end{eqnarray}

Hereafter we will omit subscripts $\pm$ except for necessary cases in order to 
make expressions simple.  Since deformation of the stellar surface from 
sphere is not so large for binary stars, we may consider the stellar 
surface as a function of $\theta$ and $\varphi$ and write it as 
$R(\theta,\varphi)$ in this coordinate. 

In the coordinate systems adopted here, we can analytically integrate the 
gravitational potential (\ref{pot}) along the $r$-direction from the center 
to the surface of the star $R(\theta,\varphi)$ because the density $\rho$ is 
constant.  For incompressible stars, we only need to solve for the shape of 
the surface $R(\theta,\varphi)$ because physical quantities such as the 
pressure or the gravitational potential are completely determined from the 
shape of the surface (see e.g. Eriguchi, Hachisu \& Sugimoto 1982; 
Eriguchi \& Hachisu 1983; Eriguchi \& Hachisu 1985).  

The stream function $\Psi$ is also expressed in terms of the surface 
variable.  Since $\zeta$ is assumed to be constant, we can explicitly 
write down the stream function $\Psi$ by using power series expansion 
on the equatorial plane as follows:  
\begin{eqnarray} \label{sol}
\lefteqn{\Psi(r\,\sin\theta,\varphi)\,=}\nonumber \\
&&-\,\zeta\biggl\{{1 \over 4}(r\,\sin\theta)^2
\,+\,\sum_{m=0}^\infty a_m\,(r\,\sin\theta)^m\,\cos m\,\varphi \biggr\} \, .
\end{eqnarray}
Imposing the boundary condition (\ref{bcon2}), i.e. 
\begin{equation} \label{bcon3}
\Psi(R(\pi/2,\varphi),\varphi)\,=\,0,
\end{equation}
on the equatorial surface, we can obtain coefficients $a_m$ for a given 
vorticity.  Components of the fluid velocity in the rotational frame 
$(u_x,u_y)$ are computed by differentiating this expression.  

Substituting the above formulae into the Bernoulli equation on the fluid 
surface and imposing the boundary condition (\ref{bcon1}) for the pressure 
there, we finally obtain the basic equations for numerical computations. 
They are basically the equations ({\ref{berno}) and (\ref{bcon3}) (see 
Appendix A).  The derived equations contain two unknown variables 
$R(\theta,\varphi)$ and $a_m$, and several unknown physical constants.  
These equations are calculated in the coordinate system $(\theta,\varphi)$ 
over the region $\theta \in [0,\pi/2]$ and $\varphi \in [0,\pi]$.  
This region is discretized equidistantly into grid points as follows: 
$(\theta_{i_\theta}, \varphi_{i_\varphi})$ 
($0 \le i_\theta \le N_\theta,\,0 \le i_\varphi \le N_\varphi$) and 
we choose $(N_\theta,N_\varphi)=(16,32)$.  
As for the coefficients $a_m$, we include terms up to $m=N_\varphi/2$.  
The trapezoidal formula is used for the integration of the gravitational 
potential.  

One model of a binary system can be numerically computed by specifying 
some physical quantities which characterize the binary system.  
We can choose the values of physically meaningful parameters 
so as to be able to compute an appropriate sequence by changing some 
of them.  For the present purpose, it is convenient to specify the mass 
ratio and the separation of two component stars as two parameters.  For 
other parameters, we choose relations for the vorticity of the fluid star 
(see Appendix B).  
Since we consider equilibrium sequences whose vorticity is conserved in the 
inertial frame as stars evolve adiabatically by the force of the gravitational 
radiation reaction potential, the following quantity must be constant in time 
as well as in space \cite{mi74}: 
\begin{equation} \label{vor}
\zeta\,+\,2\,\Omega\,=\,\zeta_0\,=\,constant, 
\end{equation} 
where $\zeta_0$ is the vorticity in the inertial frame of reference.  
In particular, the stars with $\zeta_0\,=\,0$ are usually referred to as 
irrotational configurations.  

The basic equations are discretized on the mesh points.  Since these 
difference equations are nonlinear equations for $R(\theta,\varphi)$,
$a_m$ and unknown constants, $C$ and $\Omega$, the Newton-Raphson method 
is applied to solve them.  Since the Newton-Raphson scheme requires a 
rather large matrix, we were able to employ only the mesh number mentioned 
above.  For typical models, our computational time was $\sim 3.5$ minutes 
for $1$ iteration for a fluid star-fluid star binary system and $\sim 5.7$ 
seconds for a point mass-fluid star binary system by using a Fujitsu VX vector 
computer.  A further discussion on the solving method is described in 
the appendices in detail.  

\subsection{Consistency at the stellar surface boundary}
%

%%%%%%%%%%%%%%%%% begin %%%%%%%%%%%%%%%%%
The formulation of the problem mentioned above is mathematically well-defined 
as a boundary value problem for the elliptic partial differential equation.  
Except for rigidly rotating models, this formulation is essentially the same 
as the classical problem of two-dimensional vortex flow in a rotating frame.  
We implicitly solve the equation of continuity by introducing the function 
$\Psi$, and hence we treat all equations consistently.  Actually the problem 
can be solved numerically by using the above prescription whose results we 
will show below.  

However, at the present time, we cannot show analytically that the stream 
function $\Psi$ on the equatorial plane is consistent with the boundary 
condition on the stellar surface.  We compute the stream function on the 
equatorial plane.  On the other hand, the stellar surface is determined by 
the condition of equation (\ref{bcon1}).  Since the flow lines correspond 
to the $\Psi=constant$ lines, these should coincide with the $p=constant$ 
lines.  In other words, since the velocity vector is always parallel to 
the stream line, the velocity vector in the rotating frame $\bbu$ and the 
normal vector of the stellar surface $\bbn$ are orthogonal everywhere on the 
surface.  In order to check our computational results, we have computed the 
following quantities on the stellar surface numerically: 
\begin{equation}
q\,=\,{\bbu \cdot \bbn \over \left| \bbu \right| \left| \bbn \right|} \ . 
\end{equation}
To be real solutions, this quantity should vanish on the stellar surface.  
The normal vector $\bbn$ can be written as follows in the spherical 
coordinates: 
\begin{equation}
\bbn \,=\, \bbe_r \,-\, 
{1\over R(\theta,\varphi)}
{\partial\,R\over\partial\,\theta}\,\bbe_\theta \,-\,
{1\over R(\theta,\varphi)\sin\theta}
{\partial\,R\over\partial\,\varphi}\,\bbe_\varphi \ ,
\end{equation}
where $\bbe_r$, $\bbe_\theta$ and $\bbe_\varphi$ are the the unit basis 
vectors of the spherical coordinates.  We have computed the quantity $q$
numerically at every grid point on the discretized stellar surface.  
Although we have carried out numerical differentiation to obtain this 
quantity, the value is always less than 0.05, i.e. $q < 0.05$ everywhere 
on the stellar surface even for highly deformed configurations.  Therefore, 
our method gives consistent solutions for stellar configurations as we expect.
One can also find an excellent coincidence of the present results with other
results which have been computed by using a totally different formulation and 
a numerical method (Ury\=u \& Eriguchi 1998a, 1998b).  This also ensures the 
validity of our present results.  
%%%%%%%%%%%%%%%%     end   %%%%%%%%%%%%%%%%%%
%
%
%

\section{Results}

It is convenient to use non-dimensional quantities to compare our results 
with those of others. Thus physical quantities are normalized by using
normalization factors as follows.

The non-dimensional orbital angular velocity $\tilde \Omega$ is defined as: 
\begin{equation} \label{ome}
{\tilde \Omega} \, = \, {\Omega \over \sqrt{G\rho}} \, .
\end{equation} 
Concerning the separation, the non-dimensional separation between centers 
of mass of two component stars ${\tilde d}_G$ is defined as
\begin{equation} \label{disg}
{\tilde d}_G\,=\,{(d_{G+}\,+\,d_{G-}) \over (3 M_{\rm tot}/4 \pi \rho)^{1/3}}
\, ,
\end{equation} 
where the total mass is expressed as $M_{\rm tot}\,=\,M_+\,+\,M_-$, and 
we define $d_{G\pm}$ as follows:
\begin{equation}
d_{G\pm}\, \equiv \,{ \rho \mid \int_V X_\pm dV \mid \over M_\pm}\, .
\end{equation}
The separation between two geometrical centers is normalized as 
\begin{equation} \label{dis}
{\tilde d}\,=\,{(d_+\,+\,d_-) \over (3 M_{\rm tot}/4 \pi \rho)^{1/3}}.
\end{equation} 
The total angular momentum $J$ and the total energy $E$ are normalized as
\begin{equation} \label{angm}
{\tilde J}\,=\,{J \over \sqrt{G}\,M_{\rm tot}^{5/3}\,\rho^{-1/6}}\, ,
\end{equation} 
and
\begin{equation} \label{ene}
{\tilde E}\,=\,{E \over G\,M_{\rm tot}^{5/3}\,\rho^{1/3}}\, .
\end{equation} 

In the Tables we also show the luminosity of gravitational radiation, which 
is related to the quadrupole moment tensors of the binary system.  
In the rotating frame of reference, the quadrupole moment tensors are 
defined as
\begin{equation}
I_{ij}^{(rot)}\,=\,\rho\int_V(x_i x_j\,-\,\delta_{ij}{\mid {\bf x} \mid^2 
\over 3})dV,
\end{equation}
where $i,j=1,2,3$ correspond to the $x,y,z$-component of the Cartesian 
coordinates, respectively.  By taking the symmetry of the system into
account, the quadrupole formula for the energy generation rate is 
written as follows:
\begin{equation}
{dE \over dt}\,=\,-{32\,G \over 5\,c^5}\,\Omega^6\,(I_{11}^{(rot)}\,-\,
I_{22}^{(rot)})^2 \, ,  
\end{equation}
where $c$ is the velocity of light.  Derivation of this formula is shown 
in the book of Misner, Thorne \& Wheeler~(1970) (see also Eriguchi, 
Futamase \& Hachisu 1990).  This energy generation rate is normalized as 
\begin{equation}
{\widetilde{dE \over dt}}\,=\, {d{\tilde E} \over d{\tilde t}}\bigg/
\bigg({v_0 \over c}\bigg)^5, 
\end{equation}
where non-dimensional time is defined as 
\begin{equation}
{\tilde t}\, = \, \sqrt{G\rho}\,t ,
\end{equation}
and 
\begin{equation}
v_0\,=\,\sqrt{GM_{\rm tot}^{2/3}\rho^{1/3}}.
\end{equation}

In the following subsections, our results for three different types of
binary configurations are discussed.  They are, 
\begin{enumerate}
\item Roche configurations with $M_+/M_- = 1$ and $0.1$ .
\item Irrotational Roche--Riemann configurations with several mass ratios.
\item Irrotational Darwin--Riemann configurations with several mass ratios.
\end{enumerate}
LRS computed several types of binary configurations by using the ellipsoidal 
approximation.  Concerning binary solutions, nothing but synchronously 
rotating fluid--fluid binary configurations with equal mass, i.e. Darwin
models, are compared in their papers with fully deformed solutions.  
We compare our results with those of LRS for models listed above and 
clarify quantitative differences.  These comparisons provide the evidence 
that our computational method works accurately.  

\subsection{Roche type binary systems}

We show our results in Tables \ref{rocheeqm}, \ref{rochem2} and 
\ref{rochem10} for Roche sequences with the mass ratio $M_+/M_-=1, 0.5$ 
and $0.1$, respectively.  These sequences can be regarded as simplified 
models of BH--NS binary systems in the limit of 
strong viscosity \cite{ko92} and in the frame of Newtonian gravity.  

In Figures \ref{figrocheeqm} and \ref{figrochem10}, we show our results and 
those of LRS1.  In Figures \ref{figrocheeqm}(a)--(c) and 
\ref{figrochem10}(a)--(c), physical quantities, 
${\tilde J}$, ${\tilde E}$ and ${\tilde \Omega}$, are plotted 
against the separation of mass centers of two component stars ${\tilde d}_G$
and against the non-dimensional angular momentum.  
The models on a sequence at the smallest separation in Figures 
\ref{figrocheeqm}(a), (b) and \ref{figrochem10}(a), (b) 
almost correspond to those at the Roche limit.  Here we define the Roche limit 
as the smallest separation of two centers of mass of two bodies. 
Note that models shown in the last entry of the Tables are not the exact 
critical models.  

As seen from these figures, two results obtained by two different methods
agree rather well even 
near the turning points of sequences. Differences of $\tilde J$ or 
$\tilde E$ between two results are less than $0.5\%$ for most part of the
sequence.  It is also shown that our values of the separation between the 
centers of mass of the fluid star and the point mass agree with those of 
LRS to within $2 \sim 6 \%$ around the models with the minimum of $\tilde J$ 
or $\tilde E$ where secular instability sets in and also around the models 
near the Roche limit stage.  Our results show that the separations where 
the instability sets in increase by the amount of roughly $\sim4\%$ and 
$\sim2\%$ for $M_+/M_-=1$ and $0.1$ sequences, respectively, and 
correspondingly the orbital frequency at the critical state decreases 
by the amount of $6\%$ and $3\%$, respectively, compared to the results of 
LRS.  

Here it should be noted that the secular instability occurs via angular 
momentum transfer between the orbital one and the spin because of the 
existence of viscosity (see e.g. Aizenman 1968; Hachisu 1979; 
Hachisu \& Eriguchi 1984a, 1984b, 1985; Lai, Rasio \& Shapiro 1993a).  
Consequently, as the separation is decreased, the system reaches first a 
state where secular instability sets in and next comes to the
Roche limit state.  This behavior is also seen in models of LRS1 for 
sequences whose mass ratios are the same as those of ours.  

\begin{table}
\caption{Equilibrium sequence of Roche type binary systems 
with equal mass}
\label{rocheeqm}
\begin{tabular}{@{}lcccccc}
$\tilde d$ & ${\tilde d}_G$ & ${\tilde \Omega}$ & $\tilde J$ & 
$\tilde E$ & $\widetilde{dE / dt}$ \cr
5.0 & 4.079 & 2.486(-1) & 4.104(-1) & -3.525(-1) & 3.883(-3) \cr
4.8 & 3.930 & 2.630(-1) & 4.039(-1) & -3.542(-1) & 4.688(-3) \cr
4.6 & 3.782 & 2.787(-1) & 3.974(-1) & -3.560(-1) & 5.693(-3) \cr
4.4 & 3.635 & 2.959(-1) & 3.909(-1) & -3.579(-1) & 6.957(-3) \cr
4.2 & 3.490 & 3.146(-1) & 3.845(-1) & -3.598(-1) & 8.556(-3) \cr
4.0 & 3.347 & 3.351(-1) & 3.782(-1) & -3.619(-1) & 1.059(-2) \cr
3.8 & 3.207 & 3.576(-1) & 3.720(-1) & -3.640(-1) & 1.319(-2) \cr
3.6 & 3.070 & 3.821(-1) & 3.661(-1) & -3.662(-1) & 1.653(-2) \cr
3.4 & 2.937 & 4.088(-1) & 3.606(-1) & -3.684(-1) & 2.083(-2) \cr
3.2 & 2.810 & 4.377(-1) & 3.555(-1) & -3.706(-1) & 2.637(-2) \cr
3.0 & 2.689 & 4.686(-1) & 3.510(-1) & -3.726(-1) & 3.347(-2) \cr
2.9 & 2.632 & 4.847(-1) & 3.491(-1) & -3.735(-1) & 3.773(-2) \cr
2.8 & 2.577 & 5.010(-1) & 3.475(-1) & -3.744(-1) & 4.249(-2) \cr
2.7 & 2.526 & 5.175(-1) & 3.461(-1) & -3.751(-1) & 4.780(-2) \cr
2.6 & 2.478 & 5.339(-1) & 3.451(-1) & -3.756(-1) & 5.366(-2) \cr
2.5 & 2.434 & 5.500(-1) & 3.446(-1) & -3.759(-1) & 6.004(-2) \cr
2.4 & 2.395 & 5.653(-1) & 3.445(-1) & -3.759(-1) & 6.685(-2) \cr
2.3 & 2.362 & 5.794(-1) & 3.450(-1) & -3.757(-1) & 7.387(-2) \cr
2.2 & 2.337 & 5.916(-1) & 3.460(-1) & -3.751(-1) & 8.076(-2) \cr
2.1 & 2.320 & 6.010(-1) & 3.477(-1) & -3.741(-1) & 8.690(-2) \cr
2.0 & 2.312 & 6.065(-1) & 3.495(-1) & -3.730(-1) & 9.131(-2) \cr
\end{tabular}
%\medskip
\end{table}
\begin{table}
\caption{Equilibrium sequence of Roche type binary systems
with $M_+/M_-$=0.5}
\label{rochem2}
\begin{tabular}{@{}lcccccc}
$\tilde d$ & ${\tilde d}_G$ & ${\tilde \Omega}$ & $\tilde J$ &
$\tilde E$ & $\widetilde{dE / dt}$ \cr
5.0 & 3.643 & 2.948(-1) & 3.422(-1) & -2.031(-1) & 5.427(-3) \cr
4.8 & 3.519 & 3.107(-1) & 3.370(-1) & -2.047(-1) & 6.472(-3) \cr
4.6 & 3.397 & 3.277(-1) & 3.318(-1) & -2.063(-1) & 7.749(-3) \cr
4.4 & 3.277 & 3.461(-1) & 3.268(-1) & -2.080(-1) & 9.312(-3) \cr
4.2 & 3.159 & 3.658(-1) & 3.218(-1) & -2.098(-1) & 1.123(-2) \cr
4.0 & 3.045 & 3.869(-1) & 3.170(-1) & -2.116(-1) & 1.359(-2) \cr
3.8 & 2.935 & 4.093(-1) & 3.125(-1) & -2.134(-1) & 1.648(-2) \cr
3.6 & 2.830 & 4.330(-1) & 3.082(-1) & -2.152(-1) & 2.000(-2) \cr
3.4 & 2.730 & 4.577(-1) & 3.044(-1) & -2.169(-1) & 2.427(-2) \cr
3.2 & 2.638 & 4.830(-1) & 3.011(-1) & -2.185(-1) & 2.934(-2) \cr
3.1 & 2.596 & 4.956(-1) & 2.997(-1) & -2.192(-1) & 3.218(-2) \cr
3.0 & 2.556 & 5.081(-1) & 2.986(-1) & -2.198(-1) & 3.521(-2) \cr
2.9 & 2.520 & 5.201(-1) & 2.976(-1) & -2.202(-1) & 3.840(-2) \cr
2.8 & 2.487 & 5.316(-1) & 2.970(-1) & -2.206(-1) & 4.170(-2) \cr
2.7 & 2.459 & 5.422(-1) & 2.967(-1) & -2.207(-1) & 4.501(-2) \cr
2.6 & 2.435 & 5.516(-1) & 2.967(-1) & -2.207(-1) & 4.823(-2) \cr
2.5 & 2.418 & 5.593(-1) & 2.972(-1) & -2.205(-1) & 5.115(-2) \cr
2.4 & 2.407 & 5.649(-1) & 2.979(-1) & -2.201(-1) & 5.352(-2) \cr
2.3 & 2.403 & 5.677(-1) & 2.989(-1) & -2.195(-1) & 5.501(-2) \cr
2.2 & 2.405 & 5.676(-1) & 2.996(-1) & -2.191(-1) & 5.527(-2) \cr
\end{tabular}
%\medskip
\end{table}
\begin{table}
\caption{Equilibrium sequence of Roche type binary systems 
with $M_+/M_-$=0.1}
\label{rochem10}
\begin{tabular}{@{}lcccccc}
$\tilde d$ & ${\tilde d}_G$ & ${\tilde \Omega}$ & $\tilde J$ & 
$\tilde E$ & $\widetilde{dE / dt}$ \cr
10.0 & 4.630 & 2.055(-1) & 1.407(-1) & -3.215(-2) & 2.239(-4) \cr
9.6 & 4.461 & 2.173(-1) & 1.382(-1) & -3.269(-2) & 2.699(-4) \cr
9.2 & 4.293 & 2.302(-1) & 1.356(-1) & -3.326(-2) & 3.271(-4) \cr
8.8 & 4.127 & 2.442(-1) & 1.330(-1) & -3.388(-2) & 3.988(-4) \cr
8.4 & 3.963 & 2.596(-1) & 1.304(-1) & -3.454(-2) & 4.890(-4) \cr
8.0 & 3.800 & 2.764(-1) & 1.278(-1) & -3.524(-2) & 6.032(-4) \cr
7.6 & 3.641 & 2.949(-1) & 1.252(-1) & -3.599(-2) & 7.485(-4) \cr
7.2 & 3.485 & 3.150(-1) & 1.226(-1) & -3.678(-2) & 9.336(-4) \cr
6.8 & 3.332 & 3.370(-1) & 1.200(-1) & -3.762(-2) & 1.170(-3) \cr
6.4 & 3.185 & 3.608(-1) & 1.175(-1) & -3.851(-2) & 1.473(-3) \cr
6.0 & 3.043 & 3.865(-1) & 1.150(-1) & -3.943(-2) & 1.857(-3) \cr
5.6 & 2.909 & 4.139(-1) & 1.127(-1) & -4.037(-2) & 2.341(-3) \cr
5.2 & 2.784 & 4.425(-1) & 1.105(-1) & -4.130(-2) & 2.937(-3) \cr
4.8 & 2.673 & 4.712(-1) & 1.086(-1) & -4.217(-2) & 3.646(-3) \cr
4.4 & 2.579 & 4.984(-1) & 1.071(-1) & -4.290(-2) & 4.433(-3) \cr
4.3 & 2.559 & 5.046(-1) & 1.068(-1) & -4.305(-2) & 4.631(-3) \cr
4.2 & 2.540 & 5.104(-1) & 1.066(-1) & -4.318(-2) & 4.827(-3) \cr
4.1 & 2.524 & 5.158(-1) & 1.064(-1) & -4.328(-2) & 5.014(-3) \cr
4.0 & 2.510 & 5.207(-1) & 1.062(-1) & -4.336(-2) & 5.192(-3) \cr
3.9 & 2.497 & 5.249(-1) & 1.061(-1) & -4.342(-2) & 5.354(-3) \cr
3.8 & 2.488 & 5.284(-1) & 1.061(-1) & -4.344(-2) & 5.495(-3) \cr
3.7 & 2.481 & 5.311(-1) & 1.061(-1) & -4.343(-2) & 5.610(-3) \cr
3.6 & 2.478 & 5.328(-1) & 1.062(-1) & -4.339(-2) & 5.693(-3) \cr
3.5 & 2.477 & 5.334(-1) & 1.063(-1) & -4.331(-2) & 5.737(-3) \cr
3.4 & 2.480 & 5.330(-1) & 1.065(-1) & -4.322(-2) & 5.740(-3) \cr
3.3 & 2.484 & 5.321(-1) & 1.066(-1) & -4.314(-2) & 5.715(-3) \cr
\end{tabular}
%\medskip
\end{table}
\begin{figure}
\vspace{15cm}
\caption{Physical quantities of a sequence of Roche type binary systems
with equal mass.  (a) Total angular momentum as a function of a binary 
separation.  (b) Total energy as a function of a binary separation. 
(c) Orbital angular velocity as a function of the total 
angular momentum.  Dashed and solid curves denote the sequence of LRS1 
and our sequence, respectively.  See text about the normalization factors 
for each quantity.}
\label{figrocheeqm}
\end{figure}
\begin{figure}
\vspace{15cm}
\caption{Same as Figure \protect\ref{figrocheeqm} but for quantities of 
a sequence of Roche type binary systems with $M_+/M_-=0.1$.}
\label{figrochem10}
\end{figure}

\subsection{Irrotational Roche--Riemann type binary systems}

As mentioned in Introduction, neutron stars may be composed of inviscid 
fluids (Kochanek 1992; Bildsten \& Cutler 1992).  For incompressible fluids,
the vorticity $\zeta_0$ is conserved during evolution due to the gravitational 
radiation reaction \cite{mi74}.  Furthermore the orbital angular velocity of 
the star at a large separation is much smaller than that at a separation of 
the stellar size. Since $\zeta_0$ is of the order of the orbital angular 
velocity at large distances, the constant in equation (\ref{vor}) can be
regarded very small and can be approximated to zero. In other words, almost 
all realistic neutron star binary systems can be approximated by irrotational 
configurations as far as viscosity is small.  

We show our results for irrotational Roche--Riemann sequences with  
$M_+/M_-=1, 0.5$ and $0.1$ in Tables \ref{irrreqm}, \ref{irrrm2} and 
\ref{irrrm10}, respectively.  In Figures \ref{figirrreqm} and 
\ref{figirrrm10}, our results and those of LRS1 are displayed for the 
sequences with $M_+/M_-=1$ and $0.1$.  For all the sequences, the 
turning points of $\tilde J$ and $\tilde E$ against the separation exist 
as seen in Figures \ref{figirrreqm} and \ref{figirrrm10}, which is 
qualitatively the same as that of the ellipsoidal approximation.  
It implies that there is a point where the dynamical instability will set in
on the sequence between large separation states and the Roche--Riemann limit 
of the sequence. Here the Roche--Riemann limit is defined as the 
generalization of the Roche limit to fluid stars with internal flows.  
Therefore, it is probable that an irrotational star around a massive point 
source will fall onto the point mass due to dynamical instability as the 
separation becomes smaller.  This behavior 
is the same as that obtained by the ellipsoidal approximation.  Under the 
ellipsoidal approximation, there always exists a dynamical instability point 
for any mass ratio for Roche--Riemann type binary systems before reaching 
the Roche--Riemann limit as the separation decreases (LRS1).  
The quantitative difference of the physical values between these two results 
near the turning point of the sequence amounts roughly to $3 \sim 5 \%$ as 
seen from Tables and Figures.  In particular, the separation of the minimum of 
$J$ and $E$ increases by $3.5\%$ and $2.5\%$ for the sequences with 
$M_+/M_-=1$ and $0.1$, respectively, and the orbital frequency decreases by
$5\%$ and $4\%$, respectively.  
\begin{table}
\caption{Equilibrium sequence of irrotational Roche--Riemann type binary 
systems with equal mass}
\label{irrreqm}
\begin{tabular}{@{}lcccccc}
$\tilde d$ & ${\tilde d}_G$ & $\tilde \Omega$ & $\tilde J$ & 
$\tilde E$ & $\widetilde{dE / dt}$ \cr
5.0 & 4.057 & 2.506(-1) & 3.969(-1) & -3.544(-1) & 3.987(-3) \cr
4.8 & 3.907 & 2.653(-1) & 3.895(-1) & -3.563(-1) & 4.826(-3) \cr
4.6 & 3.757 & 2.814(-1) & 3.821(-1) & -3.583(-1) & 5.877(-3) \cr
4.4 & 3.609 & 2.990(-1) & 3.746(-1) & -3.605(-1) & 7.204(-3) \cr
4.2 & 3.463 & 3.182(-1) & 3.670(-1) & -3.629(-1) & 8.886(-3) \cr
4.0 & 3.319 & 3.393(-1) & 3.595(-1) & -3.653(-1) & 1.103(-2) \cr
3.8 & 3.179 & 3.623(-1) & 3.520(-1) & -3.680(-1) & 1.378(-2) \cr
3.6 & 3.042 & 3.874(-1) & 3.447(-1) & -3.707(-1) & 1.731(-2) \cr
3.4 & 2.910 & 4.145(-1) & 3.376(-1) & -3.736(-1) & 2.183(-2) \cr
3.2 & 2.786 & 4.434(-1) & 3.309(-1) & -3.765(-1) & 2.758(-2) \cr
3.0 & 2.670 & 4.737(-1) & 3.250(-1) & -3.792(-1) & 3.481(-2) \cr
2.9 & 2.617 & 4.890(-1) & 3.223(-1) & -3.805(-1) & 3.902(-2) \cr
2.8 & 2.567 & 5.042(-1) & 3.200(-1) & -3.817(-1) & 4.361(-2) \cr
2.7 & 2.521 & 5.192(-1) & 3.181(-1) & -3.827(-1) & 4.860(-2) \cr
2.6 & 2.480 & 5.334(-1) & 3.166(-1) & -3.835(-1) & 5.384(-2) \cr
2.5 & 2.445 & 5.467(-1) & 3.156(-1) & -3.840(-1) & 5.927(-2) \cr
2.4 & 2.415 & 5.587(-1) & 3.153(-1) & -3.842(-1) & 6.475(-2) \cr
2.3 & 2.391 & 5.691(-1) & 3.156(-1) & -3.840(-1) & 7.008(-2) \cr
2.2 & 2.375 & 5.774(-1) & 3.165(-1) & -3.835(-1) & 7.491(-2) \cr
2.1 & 2.366 & 5.830(-1) & 3.180(-1) & -3.827(-1) & 7.874(-2) \cr
2.0 & 2.364 & 5.853(-1) & 3.195(-1) & -3.818(-1) & 8.086(-2) \cr
\end{tabular}
%\medskip
\end{table}
\begin{table}
\caption{Equilibrium sequence of irrotational Roche--Riemann type binary
systems with $M_+/M_-$=0.5}
\label{irrrm2}
\begin{tabular}{@{}lcccccc}
$\tilde d$ & ${\tilde d}_G$ & $\tilde \Omega$ & $\tilde J$ &
$\tilde E$ & $\widetilde{dE / dt}$ \cr
5.0 & 3.619 & 2.977(-1) & 3.334(-1) & -2.046(-1) & 5.607(-3) \cr
4.8 & 3.495 & 3.139(-1) & 3.277(-1) & -2.063(-1) & 6.700(-3) \cr
4.6 & 3.372 & 3.313(-1) & 3.220(-1) & -2.081(-1) & 8.037(-3) \cr
4.4 & 3.252 & 3.500(-1) & 3.164(-1) & -2.101(-1) & 9.674(-3) \cr
4.2 & 3.135 & 3.700(-1) & 3.109(-1) & -2.121(-1) & 1.168(-2) \cr
4.0 & 3.022 & 3.913(-1) & 3.055(-1) & -2.141(-1) & 1.413(-2) \cr
3.8 & 2.914 & 4.137(-1) & 3.004(-1) & -2.162(-1) & 1.710(-2) \cr
3.6 & 2.812 & 4.371(-1) & 2.956(-1) & -2.183(-1) & 2.067(-2) \cr
3.4 & 2.719 & 4.608(-1) & 2.913(-1) & -2.202(-1) & 2.488(-2) \cr
3.2 & 2.634 & 4.843(-1) & 2.876(-1) & -2.219(-1) & 2.971(-2) \cr
3.0 & 2.563 & 5.064(-1) & 2.849(-1) & -2.233(-1) & 3.499(-2) \cr
2.9 & 2.533 & 5.165(-1) & 2.840(-1) & -2.238(-1) & 3.769(-2) \cr
2.8 & 2.511 & 5.240(-1) & 2.835(-1) & -2.240(-1) & 3.984(-2) \cr
2.7 & 2.492 & 5.312(-1) & 2.833(-1) & -2.242(-1) & 4.208(-2) \cr
2.6 & 2.473 & 5.385(-1) & 2.833(-1) & -2.241(-1) & 4.452(-2) \cr
2.5 & 2.458 & 5.457(-1) & 2.839(-1) & -2.238(-1) & 4.725(-2) \cr
2.4 & 2.452 & 5.489(-1) & 2.848(-1) & -2.234(-1) & 4.869(-2) \cr
2.3 & 2.453 & 5.499(-1) & 2.856(-1) & -2.229(-1) & 4.934(-2) \cr
\end{tabular}
%\medskip
\end{table}
\begin{table}
\caption{Equilibrium sequence of irrotational Roche--Riemann type binary 
systems with $M_+/M_-$=0.1}
\label{irrrm10}
\begin{tabular}{@{}lcccccc}
$\tilde d$ & ${\tilde d}_G$ & $\tilde \Omega$ & $\tilde J$ & 
$\tilde E$ & $\widetilde{dE / dt}$ \cr
10.0 & 4.614 & 2.066(-1) & 1.398(-1) & -3.226(-2) & 2.280(-4) \cr
9.6 & 4.444 & 2.186(-1) & 1.373(-1) & -3.282(-2) & 2.752(-4) \cr
9.2 & 4.275 & 2.316(-1) & 1.346(-1) & -3.341(-2) & 3.341(-4) \cr
8.8 & 4.108 & 2.460(-1) & 1.320(-1) & -3.405(-2) & 4.082(-4) \cr
8.4 & 3.943 & 2.616(-1) & 1.293(-1) & -3.473(-2) & 5.015(-4) \cr
8.0 & 3.780 & 2.787(-1) & 1.266(-1) & -3.545(-2) & 6.200(-4) \cr
7.6 & 3.620 & 2.975(-1) & 1.239(-1) & -3.623(-2) & 7.707(-4) \cr
7.2 & 3.463 & 3.179(-1) & 1.213(-1) & -3.706(-2) & 9.631(-4) \cr
6.8 & 3.311 & 3.402(-1) & 1.186(-1) & -3.793(-2) & 1.208(-3) \cr
6.4 & 3.165 & 3.642(-1) & 1.160(-1) & -3.885(-2) & 1.520(-3) \cr
6.0 & 3.026 & 3.898(-1) & 1.135(-1) & -3.980(-2) & 1.912(-3) \cr
5.6 & 2.896 & 4.167(-1) & 1.111(-1) & -4.076(-2) & 2.395(-3) \cr
5.2 & 2.779 & 4.439(-1) & 1.090(-1) & -4.168(-2) & 2.973(-3) \cr
4.8 & 2.677 & 4.703(-1) & 1.072(-1) & -4.250(-2) & 3.627(-3) \cr
4.4 & 2.596 & 4.937(-1) & 1.059(-1) & -4.314(-2) & 4.304(-3) \cr
4.3 & 2.579 & 4.987(-1) & 1.057(-1) & -4.326(-2) & 4.465(-3) \cr
4.2 & 2.565 & 5.033(-1) & 1.055(-1) & -4.336(-2) & 4.616(-3) \cr
4.1 & 2.552 & 5.074(-1) & 1.053(-1) & -4.343(-2) & 4.760(-3) \cr
4.0 & 2.542 & 5.110(-1) & 1.052(-1) & -4.347(-2) & 4.888(-3) \cr
3.9 & 2.533 & 5.140(-1) & 1.052(-1) & -4.349(-2) & 5.001(-3) \cr
3.8 & 2.528 & 5.162(-1) & 1.052(-1) & -4.349(-2) & 5.092(-3) \cr
3.7 & 2.524 & 5.177(-1) & 1.053(-1) & -4.345(-2) & 5.158(-3) \cr
3.6 & 2.523 & 5.184(-1) & 1.054(-1) & -4.339(-2) & 5.199(-3) \cr
\end{tabular}
%\medskip
\end{table}
\begin{figure}
\vspace{15cm}
\caption{Same as Figure \protect\ref{figrocheeqm} but for quantities of 
a sequence of irrotational Roche--Riemann type binary systems 
with equal mass.}
\label{figirrreqm}
\end{figure}
\begin{figure}
\vspace{15cm}
\caption{Same as Figure \protect\ref{figirrreqm} but for $M_+/M_-=0.1$.}
\label{figirrrm10}
\end{figure}

\subsection{Irrotational Darwin--Riemann type binary systems}

In LRS2 they calculated physical quantities for irrotational Darwin--Riemann 
sequences by assuming ellipsoidal configurations.  As shown in the previous 
sections, exact computations have given qualitatively the same results for
models even with smaller separations. Thus we have further computed 
stationary solutions of two fluid stars with $\zeta_0=0$.  
These types of binary systems can be regarded as models of viscosity-free 
NS--NS binary systems just before coalescence.  Our numerically exact 
results for an irrotational Darwin--Riemann sequence with equal mass are 
tabulated in Table \ref{irdreqm}.  In Figures \ref{figirdreqm}(a)--(c) 
we show our results and those of LRS2.  Two results 
agree with each other very well even near the turning point of the sequence 
to within $0.5\%$.  Two fluid components of our present computation come to 
contact each other at the point of the smallest separation in 
Figure~\ref{figirdreqm} and Table~\ref{irdreqm}.  

The turning point of $\tilde J$ or $\tilde E$ corresponds to the dynamical 
instability point of the sequence just as the Roche--Riemann binary systems.  
The dynamical instability point appears before the two stars come to contact 
as the separation decreases.  This suggests that the binary configuration 
becomes dynamically unstable when the separation decreases due to 
gravitational radiation emission before these two fluid components contact 
each other.  This behavior of the sequence is the same as that obtained
in LRS2.  The separation of the sequence where $J$ or $E$ attains its 
minimum is larger by $6\%$ and the corresponding frequency is  
$10\%$ smaller than those of LRS's result.  These differences from those 
of LRS are larger than those for the point mass-fluid star binary systems.  
It is because, for the fluid star-fluid star binary system, both components 
are subject to tidal effect from the other companion star.  

\begin{table}
\caption{Equilibrium sequence of irrotational Darwin--Riemann type 
binary systems with equal mass
}
\label{irdreqm}
\begin{tabular}{@{}lcccccc}
$\tilde d$ & ${\tilde d}_G$ & $\tilde \Omega$ & $\tilde J$
 & $\tilde E$ & $\widetilde{dE / dt}$ \cr
5.0 & 4.060 & 2.506(-1) & 3.973(-1) & -6.589(-1) & 4.002(-3) \cr
4.8 & 3.910 & 2.653(-1) & 3.900(-1) & -6.608(-1) & 4.848(-3) \cr
4.6 & 3.761 & 2.814(-1) & 3.826(-1) & -6.628(-1) & 5.911(-3) \cr
4.4 & 3.613 & 2.989(-1) & 3.753(-1) & -6.649(-1) & 7.254(-3) \cr
4.2 & 3.468 & 3.181(-1) & 3.679(-1) & -6.672(-1) & 8.965(-3) \cr
4.0 & 3.326 & 3.391(-1) & 3.606(-1) & -6.696(-1) & 1.115(-2) \cr
3.8 & 3.187 & 3.620(-1) & 3.535(-1) & -6.721(-1) & 1.397(-2) \cr
3.6 & 3.053 & 3.869(-1) & 3.466(-1) & -6.747(-1) & 1.761(-2) \cr
3.4 & 2.925 & 4.138(-1) & 3.402(-1) & -6.773(-1) & 2.230(-2) \cr
3.2 & 2.805 & 4.422(-1) & 3.345(-1) & -6.798(-1) & 2.833(-2) \cr
3.1 & 2.749 & 4.568(-1) & 3.320(-1) & -6.809(-1) & 3.194(-2) \cr
3.0 & 2.696 & 4.714(-1) & 3.298(-1) & -6.819(-1) & 3.597(-2) \cr
2.9 & 2.648 & 4.861(-1) & 3.281(-1) & -6.828(-1) & 4.045(-2) \cr
2.8 & 2.604 & 5.002(-1) & 3.270(-1) & -6.834(-1) & 4.533(-2) \cr
2.7 & 2.565 & 5.138(-1) & 3.264(-1) & -6.837(-1) & 5.058(-2) \cr
2.6 & 2.533 & 5.264(-1) & 3.265(-1) & -6.836(-1) & 5.616(-2) \cr
2.5 & 2.505 & 5.385(-1) & 3.275(-1) & -6.831(-1) & 6.228(-2) \cr
2.4 & 2.489 & 5.468(-1) & 3.291(-1) & -6.823(-1) & 6.716(-2) \cr
2.3 & 2.478 & 5.538(-1) & 3.315(-1) & -6.810(-1) & 7.202(-2) \cr
2.2 & 2.475 & 5.581(-1) & 3.341(-1) & -6.795(-1) & 7.573(-2) \cr
2.1 & 2.478 & 5.596(-1) & 3.365(-1) & -6.782(-1) & 7.775(-2) \cr
2.0 & 2.482 & 5.588(-1) & 3.371(-1) & -6.778(-1) & 7.755(-2) \cr
\end{tabular}
%\medskip
\end{table}
\begin{figure}
\vspace{15cm}
\caption{Same as Figure \protect\ref{figrocheeqm} but for quantities of 
a sequence of irrotational Darwin-Riemann type binary systems with equal 
mass.}
\label{figirdreqm}
\end{figure}

In Table \ref{irdrm2}, we tabulate a sequence of irrotational fluid--fluid 
binary configurations with $M_+/M_-=0.5$.  Physical quantities of this 
sequence are also plotted in Figure \ref{figirdrm2}.  By considering that 
the mass of neutron stars ranges from the minimum mass of 
$\sim 0.7 M_{\sun} $ to the maximum mass of $\sim 1.4 M_{\sun}$, it is 
reasonable to take the minimum of the mass ratio to $M_+/M_-=0.5$.  
For any mass ratio, our results coincide qualitatively with those of 
LRS again, except for the sequences of the mass ratios around $M_+/M_-=0.1$. 
Since the primary star deforms considerably for models on such
sequences, results of our computations become less 
accurate as far as we employ the grid numbers mentioned before.
\begin{table}
\caption{Equilibrium sequence of irrotational Darwin--Riemann type binary 
systems with $M_+/M_-$=0.5
}
\label{irdrm2}
\begin{tabular}{@{}lcccccc}
$\tilde d$ & ${\tilde d}_G$ & $\tilde \Omega$ & $\tilde J$ & 
$\tilde E$ & $\widetilde{dE / dt}$ \cr
5.0 & 4.420 & 2.205(-1) & 3.683(-1) & -6.875(-1) & 2.056(-3) \cr
4.8 & 4.250 & 2.339(-1) & 3.612(-1) & -6.892(-1) & 2.506(-3) \cr
4.6 & 4.081 & 2.486(-1) & 3.540(-1) & -6.909(-1) & 3.078(-3) \cr
4.4 & 3.913 & 2.650(-1) & 3.467(-1) & -6.928(-1) & 3.812(-3) \cr
4.2 & 3.746 & 2.830(-1) & 3.394(-1) & -6.948(-1) & 4.761(-3) \cr
4.0 & 3.580 & 3.031(-1) & 3.320(-1) & -6.970(-1) & 6.002(-3) \cr
3.8 & 3.417 & 3.254(-1) & 3.246(-1) & -6.993(-1) & 7.640(-3) \cr
3.6 & 3.256 & 3.502(-1) & 3.172(-1) & -7.018(-1) & 9.825(-3) \cr
3.4 & 3.099 & 3.778(-1) & 3.100(-1) & -7.044(-1) & 1.277(-2) \cr
3.2 & 2.948 & 4.083(-1) & 3.031(-1) & -7.071(-1) & 1.677(-2) \cr
3.0 & 2.804 & 4.417(-1) & 2.969(-1) & -7.098(-1) & 2.222(-2) \cr
2.8 & 2.674 & 4.773(-1) & 2.919(-1) & -7.121(-1) & 2.963(-2) \cr
2.75 & 2.644 & 4.864(-1) & 2.910(-1) & -7.126(-1) & 3.186(-2) \cr
2.7 & 2.615 & 4.953(-1) & 2.902(-1) & -7.130(-1) & 3.422(-2) \cr
2.65 & 2.589 & 5.042(-1) & 2.897(-1) & -7.133(-1) & 3.676(-2) \cr
2.6 & 2.565 & 5.128(-1) & 2.894(-1) & -7.134(-1) & 3.938(-2) \cr
2.55 & 2.543 & 5.211(-1) & 2.894(-1) & -7.134(-1) & 4.218(-2) \cr
2.5 & 2.525 & 5.285(-1) & 2.898(-1) & -7.132(-1) & 4.495(-2) \cr
2.45 & 2.511 & 5.351(-1) & 2.907(-1) & -7.128(-1) & 4.769(-2) \cr
2.4 & 2.503 & 5.404(-1) & 2.924(-1) & -7.119(-1) & 5.036(-2) \cr
\end{tabular}
%\medskip
\end{table}
%
%
%\begin{table}
%\caption{Equilibrium sequence of irrotational Darwin--Riemann type binary 
%systems with $M_+/M_-$=0.1
%}
%\label{irdrm10}
%\begin{tabular}{@{}lcccccc}
%$\tilde d$ & ${\tilde d}_G$ & $\tilde \Omega$ & $\tilde J$ & 
%$\tilde E$ & $\widetilde{dE / dt}$ \cr
%\end{tabular}
%%%%\medskip
%\end{table}
%
%
%
\begin{figure}
\vspace{15cm}
\caption{Same as Figure \protect\ref{figirdreqm} but for $M_+/M_-=0.5$.
}
\label{figirdrm2}
\end{figure}

\section{Discussion and conclusion}

Since we have succeeded in developing a highly accurate numerical code, 
we can investigate fully deformed configurations for various 
types of incompressible fluid binary systems as far as configurations with 
the $z$-independent planar flows are concerned.  
Among them we have concentrated on sequences of the Roche, the irrotational 
Roche--Riemann and the irrotational Darwin--Riemann type configurations in 
this paper because these sequences are physically important for the evolution 
of BH--NS or NS--NS binary systems due to gravitational radiation emission.
Concerning general binary systems for which the vorticity seen from the 
inertial frame does not vanish and the internal motion exists in the rotating
frame, such as Roche--Riemann or Darwin--Riemann binaries, we have computed 
several sequences.  However, if the initial {\it spin} of the fluid star is 
rather small compared with the final {\it orbital angular velocity}, the 
spin does not affect the final results appreciably.  Therefore we did not 
show our results for such sequences in this paper.

Differences between our results and those of LRS become relatively 
large when the two component stars are coming closer and the mass 
ratio of two components differs from unity.  
It is because stars deform considerably around the Roche lobe overflowing 
stage which cannot be expressed by the ellipsoidal approximation used 
by LRS.  However the qualitative character of the equilibrium sequences 
are the same.  In other words, the dynamical and/or secular instability point 
appears before the models reach the Roche lobe overflowing stage, as shown 
in LRS1 and LRS3.  Our results show that generally the separation at the
onset of the instability is larger and correspondingly the orbital frequency 
at that point is smaller compared to the results of the ellipsoidal 
approximation by the amount of $2\sim10\%$.  

In order to clarify the final fate of binary systems, it would be better to 
define four critical distances as follows (see e.g. Lai, Rasio \& Shapiro
1993b).
(i) $r_R$: the Roche~(Darwin)--Riemann limit below which no equilibrium
states exist, i.e. the minimum separation of binary stars; 
(ii) $r_C$: the contact limit where two stars contact each other 
(note that this limit exists only for a fluid-fluid star type binary system
with equal mass and the equilibrium sequence may continue to a Dumbbell 
type sequence); 
(iii) $r_m$: the hydrodynamical stability limit below which
stars fall down onto each other dynamically due to the tidal potential; 
and 
(iv) $r_{GR}$: the general relativistic stability limit where 
no stable orbits exist due to the effect of strong gravity.  
Although the last critical distance $r_{GR}$ does not appear in our 
present treatment, the binary stars fall down onto each other when
two stars come within $r_{GR}\sim 6 GM_{\rm tot}/c^2$.  

Concerning a typical BH--NS binary system whose masses are around 
$\sim 10\,M_{\sun}$ (BH) and $\sim 1\,M_{\sun}$ (NS), we should consider 
the general relativistic effect of BH at the final stage of evolution.  
Taniguchi \& Nakamura \shortcite{tn96} have computed the incompressible 
ellipsoidal model of an irrotational neutron star around a black hole 
which is represented by using the improved pseudo-Newtonian potential.  
In their paper they studied models for which relativistic effect dominates, 
i.e. the models with $r_R<r_m<r_{GR}$ as typical BH--NS systems. 
It implies that such binary systems become dynamically unstable due to
the general relativistic effect but not due to the tidal effect 
and that the dynamical instability occurs before the star deforms 
significantly.  

Since the point of dynamical instability always appears prior to the
Roche~(Darwin)--Riemann overflowing state as shown by LRS as far as the
ellipsoidal approximation is concerned, a possibility of 
occurrence of the Roche~(Darwin)--Riemann overflow has not been considered 
seriously.  Therefore their conclusion is simple, i.e. that 
binary systems necessarily result in dynamical coalescence, although 
general relativity, tidal effect and compressibility may change the results.  
On the other hand, for {\it realistic} BH--NS binary systems, there may 
exist configurations for which the Roche--Riemann overflow will occur prior to 
dynamically falling as an alternative final fate, i.e. models  
with $r_m<r_R$. In order to show whether such a situation can be realized, 
we should take the {\it compressibility} of the fluid component into 
consideration because we have shown that there always exist dynamical 
instability points on the {\it incompressible} BH-NS binary sequences.  
We can expect that the effect of compressibility leads the fluid star to 
reach the Roche--Riemann overflow state before getting to a dynamically 
unstable state.  It is opposite to the effect of strong gravity of BH.  
This effect of compressibility will be discussed in the subsequent paper
%%%%%%%%%%%%%%%%%%%%%  begin %%%%%%%%%%%%%%%%%%%%
(Ury\=u \& Eriguchi 1998a).
%%%%%%%%%%%%%%%%%%%%%  end %%%%%%%%%%%%%%%%%%%%

If the equilibrium sequence of a binary system cannot reach a point where 
dynamical instability sets in, we need to consider the problem whether this 
Roche--Riemann overflow proceeds violently or not.  It will certainly depend 
on the compressibility and the gravitational field of the black hole.
Roughly speaking, for the fluids with stiff equations of state, if the star 
loses mass, the radius of the star is likely to shrink. Thus the mass overflow
will stop. However, since there is gravitational wave emission, the system 
will evolve to the Roche--Riemann overflow stage again. In this case the
mass overflow will continue on the time scale of gravitational wave emission.
On the other hand, if the equation of state is soft, the radius of a mass 
losing star will expand. It implies that the mass overflow will continue 
violently.  Such a mass transfer between two stars has been discussed by many 
authors (see Kochanek 1992; Bildsten \& Cutler 1992 and references therein).
When we consider the black hole, this situation may be analogous 
to the runaway mass overflow for self-gravitating axisymmetric toroidal 
configurations (see e.g. Nishida et al. 1996; Masuda \& Eriguchi 1997).

Although we have only considered incompressible fluids in this paper, the
compressibility would make Roche~(Darwin)--Riemann overflow more probable.  
Consequently the final fate of real binary systems with non-equal masses 
could be determined from the competition between the occurrence of the 
state of the marginally stable orbit due to general relativity and the 
occurrence of the state of Roche~(Darwin)--Riemann overflow.  For 
compressible gases, we have recently succeeded in developing a new 
computational method for irrotational gaseous binary systems with equal 
mass \cite{ue98a}.  We will be able to investigate more realistic situations 
in the near future as long as Newtonian gravity is concerned, and the 
results presented here will be extensively used to calibrate the results 
computed by the new computational code.  Concerning the distance $r_{GR}$, 
we will be able to implement the general relativistic effect in our 
treatment of the Roche--Riemann type binary systems by using the 
pseudo-Newtonian potential as was done by Taniguchi \& Nakamura 
\shortcite{tn96}.  After such computations, the nature of 
the final fate will be finally clarified.  

\section*{Acknowledgments}

One of us (KU) would like to thank Profs. Dennis W. Sciama and John C. 
Miller and Dr. Antonio Lanza for their warm hospitality at ICTP and SISSA.
He would also like to thank Prof. Marek Abramowicz and Dr. Vladimir Karas 
for their encouragements.  
A part of the numerical computation was carried out at the Astronomical Data 
Analysis Center of the National Astronomical Observatory, Japan.  
\appendix

\section{Basic equations for actual computations}

In this section, we describe the basic equations used in our actual 
computations.  To clarify the discussion hereafter, we call the 
stellar surface $R(\theta,\varphi)$ and Fourier coefficients $a_m$ as 
the physical variables or the variables and the other quantities, 
$\zeta, \Omega, d$ and $C$ as the physical parameters or 
the constants.  They are the unknowns which appear in our 
formulation as will be seen below.  

Because of the incompressibility and the choice of the coordinate mentioned 
in section 2, we can integrate the gravitational potential of the 
fluid and express it in terms of the stellar surface $R(\theta,\varphi)$.  
It means that the value of the gravitational potential of the 
incompressible fluid is determined only from the shape of the stellar 
surface $R(\theta,\varphi)$.  Consequently we can reduce the physical 
variables in the Bernoulli equation (\ref{berno}) to the stellar surface 
$R(\theta,\varphi)$, the Fourier coefficients $a_m$ and other constants.  
Therefore we will describe each term of this Bernoulli equation on 
the surface.  

The first term of the LHS of equation (\ref{berno}) on the surface is 
calculated after substituting equation (\ref{sol}) into (\ref{strea}) as 
follows: 
\begin{eqnarray} \label{apenet}
\lefteqn{{1 \over 2}(u_x^2 + u_y^2)\,=\,
{1 \over 2}\zeta^2\biggl\{{1 \over 4}\left[R(\theta,\varphi)\,
\sin\theta\right]^2\,\biggr.} \nonumber \\
&&+\,\sum_{m=1}^\infty m\,a_m\,\left[R(\theta,\varphi)\,
\sin\theta\right]^m\,\cos m\,\varphi \, \nonumber \\
&&+\,\left[\sum_{m=1}^\infty m\,a_m\,\left[R(\theta,\varphi)\,
\sin\theta\right]^{m-1} \,\sin m\,\varphi \right]^2 \, \nonumber \\
&&+\biggl.\,\left[\sum_{m=1}^\infty m\,a_m\,\left[R(\theta,\varphi)\,
\sin\theta\right]^{m-1} \,\cos m\,\varphi \right]^2 \biggr\}. \,
\end{eqnarray}

The second term of the LHS of equation (\ref{berno}) on the surface is 
directly expressed from equation (\ref{sol}) as 
\begin{eqnarray} \label{apsol}
\lefteqn{(\zeta\,+\,2\Omega)\,\Psi\,=\,-(\zeta\,+\,2\Omega)\,\zeta 
\biggl\{{1 \over 4} \left[R(\theta,\varphi)\,
\sin\theta\right]^2 \biggr. } \nonumber \\
&&\qquad\quad\biggl. +\,\sum_{m=0}^\infty a_m \, \left[R(\theta,\varphi)\,
\sin\theta\right]^m\,\cos m\,\varphi \biggr\} \, .
\end{eqnarray}
For the fluid star-fluid star binary system, we need to add subscripts 
$\pm$ in the two equations shown above.  
The third term of the LHS of equation (\ref{berno}) on the surface is 
calculated as follows for the fluid star-fluid star binary: 
\begin{eqnarray} \label{apcent}
\lefteqn{-{1 \over 2}\Omega^2(X_\pm^2+Y_\pm^2)\,=\, 
-{1 \over 2}\Omega^2 \left\{ \left[ R_\pm(\theta_\pm,\varphi_\pm) \,
\sin\theta_\pm \right]^2 \right.\,} \nonumber \\
&&\qquad\left. \pm\, 2 d_\pm\,R_\pm(\theta_\pm,\varphi_\pm)\,
\sin\theta_\pm\,\cos\varphi_\pm
\,+\,d_\pm^2 \right\} \,.
\end{eqnarray}
The fourth term of the LHS vanishes identically on the stellar surface.  

The fifth term of the LHS of equation (\ref{berno}) which is the 
gravitational potential term is divided into two parts: 
1) the contribution from its own component $\phi_s$ and 2) that from 
the other $\phi_e$, i.e. the relation $\phi = \phi_s+\phi_e$ holds.  
These gravitational potentials are expressed in the integral form 
of the Poisson equation (see equation (\ref{pot})). Since the Green's 
function can be expanded in terms of the Legendre functions, we can 
analytically integrate the potentials along the radial direction (see e.g. 
Eriguchi, Hachisu \& Sugimoto 1982; Eriguchi \& Hachisu 1983; Eriguchi \& 
Hachisu 1985).  After the integration, the gravitational potentials on the 
stellar surface can be expressed as follows.  The self-gravity of its own 
component becomes 
\begin{eqnarray} \label{apsfgp}
\lefteqn{\phi_s(\theta,\varphi) = - \,G \rho\int_0^\pi 
\sin\theta'\,d\theta'\,\int_0^{2\pi}d\varphi'\,}\nonumber \\
&&\qquad\qquad \times \sum_{n=0}^\infty
f_n[R(\theta,\varphi),R(\theta',\varphi')]\,P_n(\cos\beta), 
\end{eqnarray}
where $f_n[R(\theta,\varphi),R(\theta',\varphi')]$ is defined as 
\begin{eqnarray}
\lefteqn{f_n[R(\theta,\varphi),R(\theta',\varphi')]  = }  \nonumber \\
&&\left\{  
\begin{array}{cc}
\lefteqn{{1 \over n+3}{R(\theta',\varphi')^{n+3} \over 
R(\theta,\varphi)^{n+1}}} \\\\ 
\lefteqn{\quad {\rm for} \quad R(\theta',\varphi') \le 
R(\theta,\varphi),} \\\\
\lefteqn{\left({1 \over n+3}+{1 \over n-2}\right)
R(\theta,\varphi)^2 }\\\\
\lefteqn{\qquad \qquad -\,{1 \over n-2}{R(\theta,\varphi)^n \over 
R(\theta',\varphi')^{n-2}}} \\\\ 
\lefteqn{\quad {\rm for} \quad R(\theta,\varphi) \le 
R(\theta',\varphi')\ {\rm and}\ n\neq 2,} \\\\
\lefteqn{R(\theta,\varphi)^2 \left[{1 \over 5}\,+
\,\ln{R(\theta,\varphi) \over  R(\theta',\varphi')}\right] } \\\\ 
\lefteqn{\quad {\rm for} \quad
R(\theta,\varphi) \le R(\theta',\varphi')\ {\rm and}\ n=2. }
\end{array} 
\right.
\end{eqnarray}
Here $P_n$ is the Legendre function and $\cos\beta$ is defined as 
\begin{equation} \label{apbeta}
\cos\beta\,=\, \cos\theta\cos\theta'+
                \sin\theta\sin\theta'\cos(\varphi-\varphi'). 
\end{equation}
In the above expression we omit subscripts $\pm$ which should be used to
distinguish the primary and the secondary stars.  

The external gravitational potential from the other component can be 
expressed as follows: 
\begin{eqnarray} \label{apexgp}
\lefteqn{\phi_{e\pm}(\theta_\pm,\varphi_\pm)\,=\,
-G \rho\int_0^\pi\sin\theta_\mp\,d\theta_\mp\,
\int_0^{2\pi}d\varphi_\mp\,}\nonumber \\
&&\qquad \times \sum_{n=0}^\infty
{1 \over n+3}{R_\mp(\theta_\mp,\varphi_\mp)^{n+3} \over
                D_\pm^{n+1}}\,P_n(\cos\gamma_\pm), 
\end{eqnarray}
where $D_\pm$ and $\cos\gamma_\pm$ are defined as 
\begin{eqnarray} \label{apdist}
\lefteqn{D_\pm\,=\,\left\{(d_++d_-)^2\,+\,
R_\pm(\theta_\pm,\varphi_\pm)^2\right.\,} \nonumber \\
&&\qquad \left.\pm\,2(d_++d_-)R_\pm(\theta_\pm,\varphi_\pm)
\sin\theta_\pm\cos\varphi_\pm\right\}^{1/2}, 
\end{eqnarray}
and 
\begin{eqnarray} \label{apgam}
\lefteqn{\cos\gamma\,=\,\left\{ \pm(d_++d_-)
R_\mp(\theta_\mp,\varphi_\mp)
\sin\theta_\mp\cos\varphi_\mp\, \right.}
\nonumber \\
&&\!\!\!\!\!\!\!+\,\left. R_\pm(\theta_\pm,\varphi_\pm)
R_\mp(\theta_\mp,\varphi_\mp)\cos\beta \right\}
/ R_\mp(\theta_\mp,\varphi_\mp) D_\pm ,
\end{eqnarray}
respectively.  Because of the symmetry of the $x$-$z$ and the $x$-$y$ 
planes, we can impose the following condition for the shape of the star: 
\begin{equation} \label{apsur}
R(\theta,\varphi)= R(\pi-\theta,\varphi)= 
R(\theta,2\pi-\varphi) .  
\end{equation}
Consequently we can reduce the integral region to a quarter of the whole 
space, 
\begin{equation} \label{apint}
\int_0^\pi d\theta'\int_0^{2\pi} d\varphi' 
\quad\longrightarrow\quad 
\int_0^{\pi/2} d\theta'\int_0^\pi d\varphi' .
\end{equation}
Accordingly, the Legendre polynomials in the gravitational potentials 
should be rearranged.  For $P_n(\cos\beta)$ in $\phi_s$, 
we should replace it as follows: 
\begin{eqnarray} \label{apleg}
\lefteqn{P_n(\cos\beta) \quad \longrightarrow \quad
P_n(\cos\beta_a)\,+\,P_n(\cos\beta_b) }\nonumber \\
&&\qquad\qquad\quad +\,P_n(\cos\beta_c)\,+\,P_n(\cos\beta_d)  ,
\end{eqnarray}
where
\begin{eqnarray}
\cos\beta_a &=& \cos\theta\cos\theta'+
   \sin\theta\sin\theta'\cos(\varphi-\varphi'), \nonumber \\
\cos\beta_b &=&- \cos\theta\cos\theta'+
   \sin\theta\sin\theta'\cos(\varphi-\varphi'), \nonumber \\
\cos\beta_c &=& \cos\theta\cos\theta'+
   \sin\theta\sin\theta'\cos(\varphi+\varphi'), \nonumber \\
\cos\beta_d &=&- \cos\theta\cos\theta'+
   \sin\theta\sin\theta'\cos(\varphi+\varphi').
\end{eqnarray}
The quantity $P_n(\cos\gamma)$ in the external potential is rearranged in 
the same way.  

As we mentioned in section 2, the Bernoulli equation is discretized on 
the surface grid points.  
%%The number of the Bernoulli equation on each 
%%grid points $(i_\theta, i_\varphi)$ i.e. $(N_\theta+1) \times (N_\varphi+1)$ 
%%is the same as the number of the discretized surface $R(\theta_{i_\theta}, 
%%\varphi_{i_\varphi})$.  
The discrete approximation to the Bernoulli equation gives one equation
for each point on the surface grid $(i_\theta, i_\varphi)$ and can 
therefore be used to determine the surface $R(\theta_{i_\theta}, 
\varphi_{i_\varphi})$.

We have another physical variable $a_m$ in our formulation.  
As mentioned in section 2, they are determined from the boundary condition 
of the stream function $\Psi$.  By using equation~(\ref{sol}), 
equation~(\ref{bcon3}) can be expressed as follows:  
\begin{equation} \label{apbcon}
{1 \over 4}R(\pi/2,\varphi)^2\,+\,\sum_{m=0}^\infty a_m 
R(\pi/2,\varphi)^m \cos\,m\varphi\,=\,0 .
\end{equation}
This equation is also discretized on the grid points of 
$\varphi_{i_\varphi}$.  Since we include the coefficients up to 
$0\le m \le N_\varphi/2$, if we adopt the grid points with even integers of 
$i_\varphi$ on the equatorial plane, the number of the equations and the
number of the coefficients coincide. As an alternative way,  it is possible 
to use all the grid points on the equatorial surface of the star and 
employ the least square method to determine $a_m$.  

\section{Conditions for parameters}

In this section we describe the scheme to determine the physical parameters 
or the constants.  The choice of the parameters is essential to construct 
a successful numerical code which can be used to get converged equilibrium 
solutions.  In our actual computations, we use non-dimensional variables
as follows:
\begin{eqnarray}
&&{\tilde R}\,=\,{R \over R_0},\qquad
{\tilde u_i}\,=\,{u_i \over \sqrt{G\rho} R_0},\qquad
{\tilde \zeta}\,=\,{\zeta \over \sqrt{G\rho}},\qquad
\nonumber \\
&&{\tilde \Omega}\,=\,{\Omega \over \sqrt{G\rho}},\qquad
{\tilde \Psi}\,=\,{\Psi \over \sqrt{G\rho}R_0^2},\qquad 
{\tilde p}\,=\,{p \over p_c},\qquad \nonumber \\
&&{\tilde C}\,=\,{C \over G \rho R_0^2},\qquad
{\tilde \phi}\,=\,{\phi \over G\rho R_0^2},\qquad
{\tilde \beta}\,=\,{p_c \over G\rho^2 R_0^2},\qquad \nonumber \\
&&{\tilde a}_m\,=\,{a_m \over R_0^{m-2}},\qquad
{\tilde X}\,=\,{X \over R_0},\qquad
{\tilde Y}\,=\,{Y \over R_0},
\end{eqnarray}
where $R_0$ is the half of the geometrical width of the primary star 
along the $x$ axis, i.e. $R_0=(R_+^{\rm out}-R_+^{\rm in})/2$ and $p_c$ 
is the pressure at the coordinate center of the each component.  

Using these non-dimensional variables, we can rewrite the Bernoulli equation 
(\ref{berno}) as follows: 
\begin{eqnarray}
\lefteqn{{1 \over 2}({\tilde u}_x^2 + {\tilde u}_y^2)\,+\,
({\tilde \zeta}\,+\,2{\tilde \Omega}){\tilde \Psi}}\nonumber \\
&&\qquad\qquad-\,{1 \over 2}{\tilde \Omega}^2
({\tilde X}^2+{\tilde Y}^2)\,+\,
{\tilde \beta}{\tilde p}\,+\,{\tilde \phi}
\,=\, {\tilde C} \, . 
\end{eqnarray}
We note that as we have shown in the previous section 
we only need to handle the equations on the stellar surface where 
the pressure vanishes.  Therefore the parameter ${\tilde \beta}$ does not 
appear in the equation but can be determined afterwards.  

For the fluid star-fluid star binary system, there are seven physical 
parameters, i.e.
\begin{equation}
{\tilde \zeta}_\pm, \quad {\tilde \Omega}, \quad {\tilde d_\pm} \quad, 
{\tilde C}_\pm.  
\end{equation}
Therefore we should impose seven conditions so as to determine these 
parameters.  As for the quantities ${\tilde \zeta}_\pm$,  we can specify 
some simple relations of ${\tilde \zeta}$ and ${\tilde \Omega}$ as 
\begin{equation}\label{aprot}
F_\pm({\tilde \zeta}_\pm,{\tilde \Omega})\,=\,0. 
\end{equation}
For the circulation preserving sequences, this relation should be
\begin{equation}
{\tilde \zeta}_\pm\,+\,{\tilde \Omega}\,=\,{\tilde \zeta}_{0\pm},   
\end{equation}
where ${\tilde \zeta}_{0\pm}$ are given constants.  For the synchronously 
rotating sequences, we can choose the relation 
\begin{equation}
\tilde{\zeta}_\pm=0 .  
\end{equation}
As for the distances from the rotational axis to the origins of the spherical
coordinate systems of each component ${\tilde d}_\pm$, two conditions are
required.  The total separation ${\tilde d}_+ + {\tilde d}_-$ and the mass 
ratio $M_+/M_-$ are specified for this purpose.  Since the origins of the 
spherical coordinates of two component stars are chosen to be located at the 
geometrical centers of the two stars on the $x$ axis, we can impose the 
following three conditions: 
%
%\begin{eqnarray}
\[
\lefteqn{{\tilde R}_+(\pi/2,0)=1,  \quad {\tilde R}_+(\pi/2,\pi)=1,}
%\nonumber \\
\]
%\begin{eqnarray}
\begin{equation}
{\tilde R}_-(\pi/2,0)={\tilde R}_-(\pi/2,\pi).
\end{equation}
%\end{eqnarray}
%
It should be noted that the physically meaningful conditions to specify the 
parameters are four, in other words, they are: 1) two conditions for 
$\zeta_\pm$, 2) one condition for the total separation and 3) one condition
for the mass ratio of the binary system.  

For the point source-fluid star binary system, the physical constants 
related to the secondary point source are different from those of the 
fluid-fluid binary system.  In this case parameters ${\tilde \zeta}_-$ 
and ${\tilde C}_-$ appear no more but we need to include one parameter 
which characterizes the gravity of the secondary star.  In this
paper we choose the mass of the secondary as a parameter:
\begin{equation}
\phi_{s+} = - {GM_- \over D_+} . 
\end{equation}
Therefore the total number of parameters becomes six for this case:  
\begin{equation}
{\tilde \zeta}_+, \quad {\tilde \Omega}, \quad {\tilde d_\pm} \quad, 
{\tilde C}_+, \quad M_-
\end{equation}
Consequently we should specify six relations for them, five of which are 
the same as the previous case: we need not use the relation for 
${\tilde \zeta}_-$ and the relation for the radius of the secondary
star.  The sixth relation is obtained from the requirement that the 
motion of the point mass should be Keplerian subjected to the 
gravitational force of the primary star.  It can be written by using 
the non-dimensional variables as follows:   
\begin{equation}
{\tilde d}_-{\tilde \Omega}^2\,=\,\sum_{n=0}^\infty
{{\tilde I}_n \over (d_++d_-)^{n+2}}, 
\end{equation}
where 
\begin{eqnarray}
\lefteqn{{\tilde I}_n \equiv {n+1 \over n+3} 
\int_0^{\pi}\sin\theta_+ d\theta_+ \int_0^{2\pi} d\varphi_+ }
\nonumber \\
&&\qquad\qquad\qquad\times{\tilde R}_+(\theta_+,\varphi_+)^{n+3} 
P_n(\cos\gamma).  
\end{eqnarray}
Therefore, the physically meaningful conditions for the parameters are three:
1) the relation for $\zeta_+$,  2) one condition for the total separation and 
3) one condition for the mass ratio of the binary system.  

These equations for the parameters are solved simultaneously with the
physical variables mentioned before by using the Newton-Raphson scheme.
This method is so powerful that we can obtain converged solutions after
only 5 or 6 iteration cycles.

\end{document}